\documentclass[
    ,final            % use final for the camera ready runs
  ]
  {aipproc}

\layoutstyle{6x9}

\usepackage{epsfig}
 \usepackage{rotating}
 \usepackage{amsmath,amssymb}
 \usepackage{graphicx}
\usepackage{ifthen,array}
 \usepackage{setspace}
 \usepackage{fancyhdr}
 \usepackage{moreverb}
 \usepackage{rotating}
\usepackage{amsfonts}
\usepackage{bbm}

\def\e6{$E(6)$}
\def\10{$SO(10)$}
\def\21{$SU(2) \otimes U(1) $}
\def\lr{$SU(2)_L \otimes SU(2)_R \otimes U(1)$}
\def\422{$SU(4) \otimes SU(2) \otimes SU(2)$}
\def\321{$SU(3) \otimes SU(2) \otimes U(1)$}
\def\lsim{\raise0.3ex\hbox{$\;<$\kern-0.75em\raise-1.1ex\hbox{$\sim\;$}}}
\def\gsim{\raise0.3ex\hbox{$\;>$\kern-0.75em\raise-1.1ex\hbox{$\sim\;$}}}
\def\lfv{lepton flavour violation }

\def\meff{\langle m_{\nu} \rangle}
\newcommand{\ed}{\end{document}}
\DeclareMathAlphabet{\mathsc}{OT1}{cmr}{m}{sc}

\newcommand{\CL}   {C.L.}
\newcommand{\dof}  {d.o.f.}

\newcommand{\eVq}  {\rm{eV}^2}
\newcommand{\Sol}  {\mathsc{sol}}
\newcommand{\Atm}  {\mathsc{atm}}

\newcommand{\Dms}  {\Delta m^2_\Sol}
\newcommand{\Dma}  {\Delta m^2_\Atm}

\def \znbb {$0\nu\beta\beta$ }

\def\meff{\langle m_{\nu} \rangle}

\let\vev\VEV
\def\ZP{$Z^\prime$ }
\def\e6{$E(6)$}
\def\10{$SO(10)$}
\def\21{$SU(2) \otimes U(1) $}
\def\lr{$SU(3) \otimes SU(2)_L \otimes SU(2)_R \otimes U(1)_{B-L}$ }
\def\422{$SU(4) \otimes SU(2) \otimes SU(2)$ }
\def\321{$SU(3) \otimes SU(2) \otimes U(1)$ }

\renewcommand{\baselinestretch}{1.05}

\newcommand{\AddrAHEP}{%
 Instituto de F\'{\i}sica Corpuscular,
  C.S.I.C. -- Universitat de Val{\`e}ncia \\
  Edificio de Institutos de Paterna, Apartado 22085,
  E--46071 Val{\`e}ncia, Spain\\}

\begin{document}
\newcommand{\Od}{{\cal O}}

\title{Neutrinos in astroparticle physics}

\author{J. W. F. Valle}{
address={\AddrAHEP}}

\begin{abstract}
  I briefly discuss the role of neutrinos as probes in astroparticle
  physics and review the status of neutrino oscillation parameters as
  of June 2006, including recent fluxes, and latest SNO, K2K and MINOS
  results.
  I comment on the origin of neutrino masses in seesaw-type and
  low-scale models and mention some of their experimental signals.
\end{abstract}
\maketitle

%%%%%%%%%%%%%%%%%%%%%%%%%%%%%%%%%%%%%%%%%%%%
%% MAINMATTER
%%%%%%%%%%%%%%%%%%%%%%%%%%%%%%%%%%%%%%%%%%%%

\bibliographystyle{aipproc}   % if natbib is available

 \section{Introduction}
\label{sec:introduction}

Neutrinos play a central role as probes in astroparticle physics and
are basic indicators of what may lie ahead of the Standard Model (SM).
The discovery of neutrino oscillations comes mainly from the study of
``heavenly''
neutrinos~\cite{fukuda:2002pe,ahmad:2002jz,Kajita:2004ga}, and has
been brilliantly confirmed by laboratory data from
reactors~\cite{araki:2004mb} and
accelerators~\cite{Ahn:2006zz,Tagg:2006sx}.

Here I summarize the status of the interpretation of the current
neutrino data within the simplest CP-conserving three-neutrino
oscillation scenario.
In addition to a determination of the solar and atmospheric
oscillation parameters $\theta_{12}$ \& $\Dms$ and $\theta_{23}$ \&
$\Dma$, one gets a constraint on the angle $\theta_{13}$.  Together
with the small ratio $\Dms/\Dma$ the angle $\theta_{13}$ holds the key
for future searches for CP violation in neutrino oscillation.  The
growing precision of oscillation experiments also opens good prospects
for improved robustness tests, probing unitarity
violation~\cite{schechter:1980gr} and other forms of non-standard
neutrino interactions.

The search for lepton number violating processes such as neutrinoless
double beta decay~\cite{elliott:2002xe,doi:1985dx} (\znbb) constitutes
a very important goal for the future, as this will probe whether
neutrinos are Dirac or Majorana particles, irrespective of the
mechanism that induces their mass. This is known as the ``black-box''
theorem~\cite{Schechter:1982bd}.  In addition, \znbb will be sensitive
to the absolute scale of neutrino mass and to CP violation induced by
the so-called Majorana phases~\cite{schechter:1980gr}, inaccessible in
conventional
oscillations~\cite{bilenky:1980cx,Schechter:1981gk,doi:1981yb}.

\section{Status of neutrino oscillations}
\label{sec:stat-neutr-oscill}

The discovery of oscillations marks a turning point in particle and
nuclear physics and implies that neutrinos have masses.  This
possibility has been first suggested by theory in the early eighties,
both on general grounds and on the basis of different versions of the
seesaw
mechanism~\cite{Minkowski:1977sc,Orloff:2005nu,Weinberg:1980bf,schechter:1980gr,schechter:1982cv,Lazarides:1980nt}.
The basic ingredient is the lepton mixing matrix, whose simplest
unitary 3-dimensional form is given as~\cite{schechter:1980gr}
\begin{equation}
  \label{eq:2227}
K =  \omega_{23} \omega_{13} \omega_{12}
\end{equation}
where each $\omega$ is effectively $2\times 2$, characterized by an
angle and a CP phase.  Majorana phases do not affect oscillations and,
moreover, current neutrino oscillation data have no sensitivity to the
remaining Dirac CP violation phase.  Thus we set the three phases to
zero.
In this approximation oscillations depend on the three mixing
parameters $\sin^2\theta_{12}, \sin^2\theta_{23}, \sin^2\theta_{13}$
and on the two mass-squared splittings $\Dms \equiv \Delta m^2_{21}
\equiv m^2_2 - m^2_1$ and $\Dma \equiv \Delta m^2_{31} \equiv m^2_3 -
m^2_1$ characterizing solar and atmospheric neutrinos.  The hierarchy
$\Dms \ll \Dma$ implies that, to a good approximation, one can set
$\Dms = 0$ in the analysis of atmospheric and accelerator data, and
$\Dma$ to infinity in the analysis of solar and reactor data.

Interpreting the data requires good calculations of the corresponding
fluxes~\cite{Bahcall:2004fg,Honda:2004yz}, neutrino cross sections and
response functions, as well as an accurate description of neutrino
propagation in the Sun and the Earth, taking into account matter
effects~\cite{mikheev:1985gs,wolfenstein:1978ue}.

The resulting three--neutrino oscillation parameters obtained in the
global analysis are summarized in Fig.~\ref{fig:global}. The analysis
includes all new neutrino oscillation data, as of June 2006, as
described in Appendix C of hep-ph/0405172 (v5)~\cite{Maltoni:2004ei}.
These include new Standard Solar Model~\cite{Bahcall:2005va}, new SNO
salt~\cite{Aharmim:2005gt}, latest K2K~\cite{Ahn:2006zz} and
MINOS~\cite{Tagg:2006sx} data.
In the upper panels of the figure the $\Delta \chi^2$ is shown as a
function of the three mixing parameters $\sin^2\theta_{12},
\sin^2\theta_{23}, \, \sin^2\theta_{13}$ and two mass squared
splittings $\Delta m^2_{21}, \Delta m^2_{31}$, minimized with respect
to the undisplayed parameters. The lower panels show two-dimensional
projections of the allowed regions in the five-dimensional parameter
space.  In addition to a confirmation of oscillations with $\Dma$,
accelerator neutrinos provide a better determination of $\Dma$ as one
can see by comparing dashed and solid lines in Fig.~\ref{fig:global}.
Clearly MINOS~\cite{Tagg:2006sx} leads to an improved determination
and a slight increase in $\Dma$.
On the other hand reactors~\cite{araki:2004mb} have played a crucial
role in selecting large-mixing-angle (LMA)
oscillations~\cite{pakvasa:2003zv} out of the previous ``zoo'' of
solutions~\cite{gonzalez-garcia:2000sq}.
\begin{figure}[t] \centering
    \includegraphics[width=.95\linewidth,height=9cm]{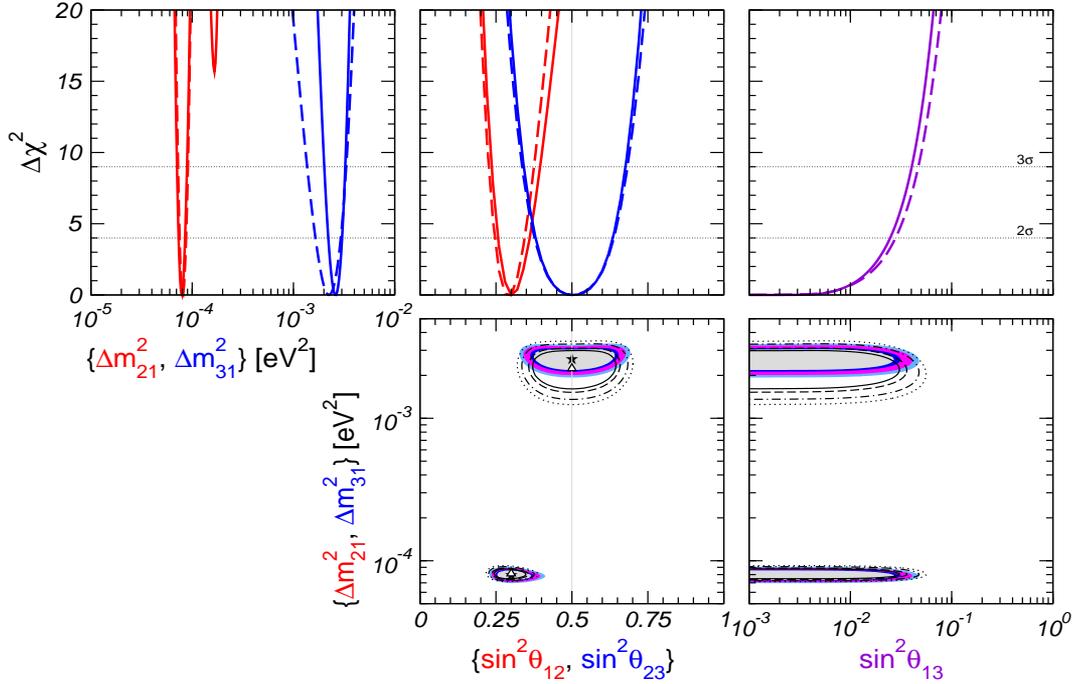}
    \caption{\label{fig:global} %
      Current neutrino oscillation regions at 90\%, 95\%, 99\%, and
      3$\sigma$ \CL\ for 2 \dof\, from Ref.~\cite{Maltoni:2004ei}. In
      top panels $\Delta \chi^2$ is minimized with respect to
      undisplayed parameters.}
\end{figure}
Table~\ref{tab:summary} summarizes the current best fit values and the
allowed 3$\sigma$ ranges that follow from the global fit.
\begin{table}[t] \centering    \catcode`?=\active \def?{\hphantom{0}}
      \begin{tabular}{|l|c|c|}        \hline        parameter & best
      fit & 3$\sigma$ range         \\  \hline\hline        $\Delta
      m^2_{21}\: [10^{-5}~\eVq]$        & 7.9?? & 7.1--8.9 \\
      $\Delta m^2_{31}\: [10^{-3}~\eVq]$ & 2.6?? &  2.0--3.2 \\
      $\sin^2\theta_{12}$        & 0.30? & 0.24--0.40 \\
      $\sin^2\theta_{23}$        & 0.50? & 0.34--0.68 \\
      $\sin^2\theta_{13}$        & 0.00 & $\leq$ 0.040 \\
      \hline
\end{tabular}    \vspace{2mm}
\caption{\label{tab:summary} Neutrino oscillation parameters as of June 2006, 
from Ref.~\cite{Maltoni:2004ei}.}
\end{table}

Note that in a three--neutrino scheme CP violation disappears when two
neutrinos become degenerate or when one of the angles
vanishes~\cite{schechter:1980bn}.  As a result CP violation is doubly
suppressed, first by $\alpha \equiv \Dms/\Dma$ and also by the small
value of $\theta_{13}$.
\begin{figure}[t]
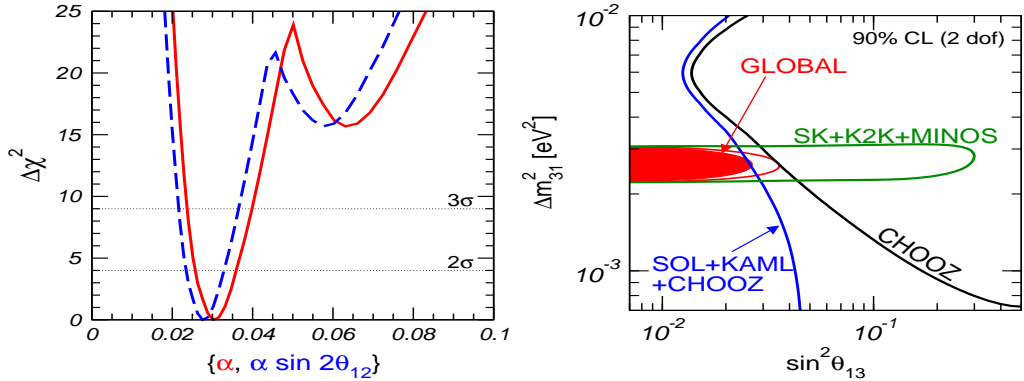
 \centering
  \includegraphics[height=5cm,width=.45\linewidth]{F-fcn.alpha06.eps}
\includegraphics[height=5cm,width=.45\linewidth]{th13-06.eps}
\caption{\label{fig:alpha}%
  $\alpha \equiv \Dms / \Dma$ and $\sin^2\theta_{13}$ bound from the
  updated analysis given in Ref.~\cite{Maltoni:2004ei}.}
\end{figure}
The left panel in Fig.~\ref{fig:alpha} gives the parameter $\alpha$,
as determined from the global $\chi^2$ analysis.
The right panel shows the impact of different data samples on
constraining $\theta_{13}$.  One sees that for larger $\Dma$ values
the bound on $\sin^2\theta_{13}$ is dominated by CHOOZ, while for low
$\Dma$ the solar and KamLAND data become quite relevant.

There is now an ambitious long-term effort towards probing CP
violation in neutrino
oscillations~\cite{Alsharoa:2002wu,apollonio:2002en,albright:2000xi}.
As a first step, upcoming reactor and accelerator long baseline
experiments aim at improving the sensitivity on
$\sin^2\theta_{13}$~\cite{Huber:2004ug}.  An alternative possibility
involving the day/night effect studies in large water Cerenkov solar
neutrino experiments such as UNO, Hyper-K or LENA has also been
suggested~\cite{Akhmedov:2004rq}.

%%% rob b %%%

Reactor neutrino data have played a crucial role in testing the
robustness of solar oscillations vis a vis astrophysical
uncertainties, such as magnetic fields in the
radiative~\cite{Burgess:2003su,Burgess:2003fj,burgess:2002we} and
convective zone~\cite{miranda:2000bi,guzzo:2001mi,barranco:2002te},
leading to stringent limits on neutrino magnetic transition
moments~\cite{Miranda:2004nz}.
KamLAND has also played a key role in identifying oscillations as
``the'' solution to the solar neutrino problem~\cite{pakvasa:2003zv}
and also in pinning down the LMA parameter region among previous wide
range of oscillation solutions~\cite{gonzalez-garcia:2000sq}.

%%% rob e %%%

\begin{figure}[t] \centering
    \includegraphics[height=3cm,width=.45\linewidth]{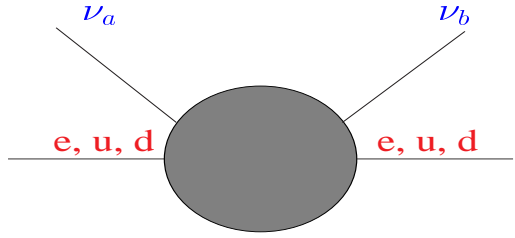}
    \caption{\label{fig:nuNSI} %
      Non-standard neutrino interactions arise, e.~g., from the
      non-unitary structure of charged current weak interactions
      characterizing seesaw-type schemes~\cite{schechter:1980gr}.}
\end{figure}

However, there still some fragility in the interpretation of the data
if sub-weak strength ($\sim \varepsilon G_F$) non-standard neutrino
interaction (NSI) operators (Fig.~\ref{fig:nuNSI}) are included.
Indeed, most neutrino mass generation mechanisms imply the existence
of such dimension-6 operators. They can be of two types:
flavour-changing (FC) and non-universal (NU). Their presence leads to
the possibility of resonant neutrino conversions even in the absence
of neutrino masses~\cite{valle:1987gv}.
While model-dependent, the expected NSI magnitudes may well
fall within the range that will be tested in future precision
studies~\cite{Huber:2004ug}.
For example, in the inverse seesaw model~\cite{Deppisch:2004fa} the
non-unitary piece of the lepton mixing matrix can be sizeable, hence
the induced non-standard interactions.  Relatively sizable NSI
strengths may also be induced in supersymmetric unified
models~\cite{hall:1986dx} and models with radiatively induced neutrino
masses~\cite{zee:1980ai,babu:1988ki}.

The determination of atmospheric neutrino parameters $\Dma$ and
$\sin^2\theta_\Atm$ is hardly affected by the presence of NSI on
down-type quarks, at least within the 2--neutrino
approximation~\cite{fornengo:2001pm}. Future neutrino factories will
substantially improve this bound~\cite{huber:2001zw}.

In contrast, the determination of solar neutrino parameters is not
quite robust against the existence of NSI~\cite{Miranda:2004nb}, even
if reactor data are included. One can show that even a small residual
non-standard interaction in this channel has dramatic consequences for
the sensitivity to $\theta_{13}$ at a neutrino
factory~\cite{huber:2001de}.  Improving the sensitivities on NSI
constitutes at a near detector or via coherent neutrino scattering off
nuclei~\cite{Barranco:2005yy} a window of opportunity for neutrino
physics in the precision age.

\section{The origin of neutrino mass}
\label{sec:origin-neutrino-mass}

Here I briefly discuss the theory of neutrino mass and mention some
recent attempts at predicting neutrino masses and mixing.

\subsection{Light Dirac neutrinos}
\label{sec:light-dirac-neutr}

Gauge theories prefer Majorana neutrinos~\cite{schechter:1980gr}.
This statement holds irrespective the detailed model of neutrino mass
generation. The emergence of Dirac neutrinos would constitute a
surprise, indicating the existence of a fundamental lepton number
symmetry whose origin should be understood. Without a specific reason,
the appearance of such symmetry would be ``accidental''.

Nevertheless there are interesting ideas for generating light Dirac
neutrinos. For example, theories involving large extra dimensions
offer a novel scenario to account for small neutrino
masses~\cite{Dienes:1998sb,Arkani-Hamed:1998vp,Faraggi:1999bm,Dvali:1999cn,Mohapatra:1999zd,Barbieri:2000mg,Ioannisian:1999sw}.
According to this picture, right-handed neutrinos propagate in the
bulk while left-handed neutrinos, being a part of the lepton doublet,
live only on the SM branes. As a result, neutrinos can naturally get
very small Dirac masses via mixing with a ``bulk'' fermion.

\subsection{Light Majorana neutrinos}
\label{sec:light-major-neutr}

\begin{figure}[h] \centering
    \includegraphics[height=3.5cm,width=.4\linewidth]{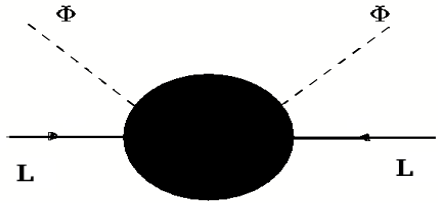}
    \caption{\label{fig:d-5} Dimension five operator responsible for
      neutrino mass~\cite{Weinberg:1980bf}.}
\end{figure}

Charged fermions in the SM come in two chiral species to provide their
mass after the electroweak symmetry breaks through the nonzero vacuum
expectation value (vev) of the Higgs scalar doublet $\vev{\Phi}$.
Neutrinos do not.  There is, however, an effective lepton number
violating dimension-five operator $\lambda L \Phi L \Phi$ in
Fig.~\ref{fig:d-5}, which can be added to the SM (here $L$ denotes any
of the lepton doublets)~\cite{Weinberg:1980bf}.
After the Higgs mechanism this induces Majorana neutrino masses
$\propto \vev{\Phi}^2$, thus providing a natural way to account for
the smallness of neutrino masses, irrespective of their specific
origin. 
Little more can be said from first principles about the \emph{
  mechanism} giving rise to this operator, its associated mass \emph{
  scale} or its \emph{flavour structure}.  Its strength $\lambda$ may
be suppressed by a large scale $M_X$ in the denominator (top-down)
scenario, leading to $ m_{\nu} = \lambda_0 \frac{\vev{\Phi}^2}{M_X}, $
where $\lambda_0$ is some unknown dimensionless constant.
Gravity has been argued to break global symmetries and thus could
induce the dimension-five operator, with $M_X = M_P$, the Planck
scale~\cite{deGouvea:2000jp}. In this case the magnitude of the
resulting Majorana neutrino masses are too small.

Alternatively, the strength $\lambda$ may be suppressed by small
parameters (e.g. scales, Yukawa couplings) and/or loop-factors
(bottom-up scenario) with no need for a large scale.
Both classes of scenarios have been reviewed in \cite{Valle:2006vb}.
Here is a brief summary.

\subsection{Seesaw-type models}
\label{sec:top-down-scenario}

\begin{figure}[ht] \centering
   \includegraphics[height=3cm,width=.38\linewidth]{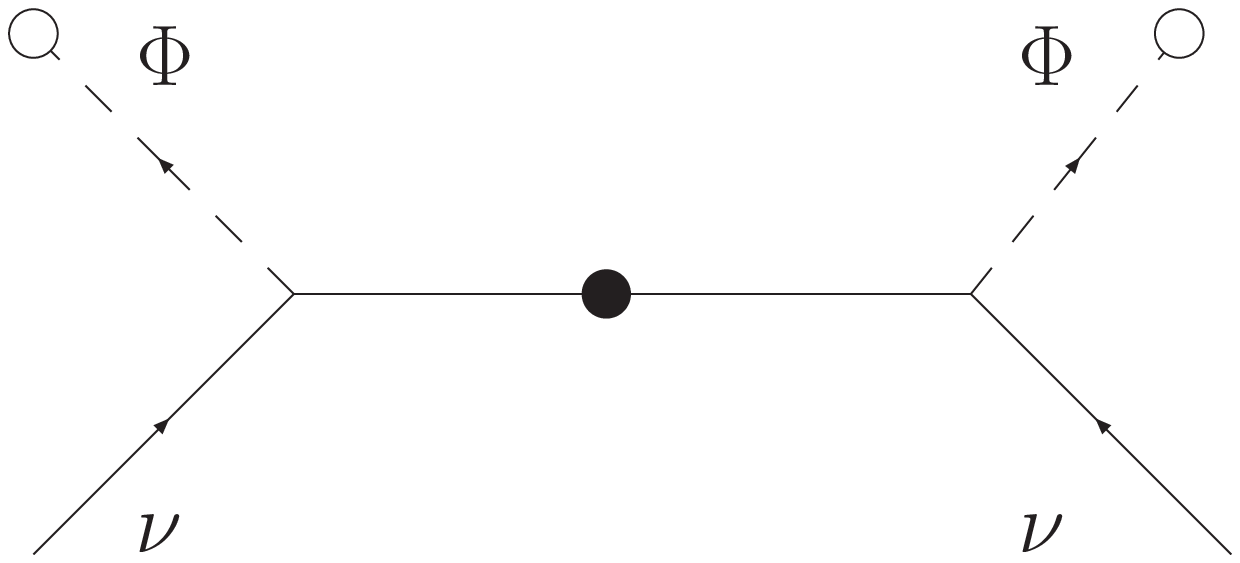} \hskip 1cm
    \includegraphics[height=3.7cm,width=.3\linewidth]{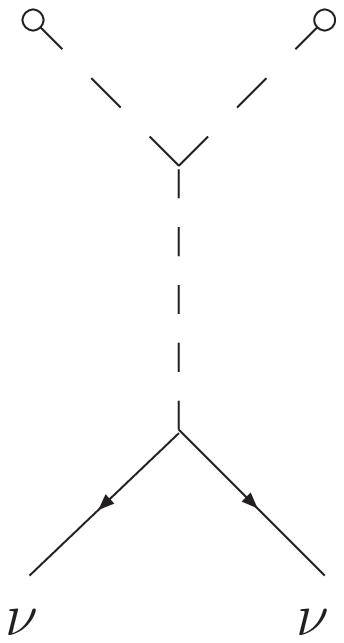}
    \caption{\label{fig:seesaw} %
      Two types of seesaw mechanism}
\end{figure}

The most popular top-down scenario is the
seesaw~\cite{Minkowski:1977sc}. The idea is to generate the dim-5
operator by the exchange of heavy states, either fermions (type-I) or
scalars (type-II), typically both, as shown in Fig.~\ref{fig:seesaw}.
The main point is that, as the masses of the intermediate states go to
infinity, neutrinos become light~\cite{Orloff:2005nu}. The seesaw
provides a simple realization of Weinberg's dim-5
operator~\cite{Weinberg:1980bf}. It can be implemented in many ways,
with explicitly or spontaneously broken B-L, gauged or not; with
different gauge groups and multiplet contents, minimal or not; with
its basic scale large or small. Seesaw with gauged B-L broken at large
scale is but one possibility.  I have no space here for a detailed
discussion, those interested in a short seesaw ``Kamasutra'' may
consult Ref.~\cite{Valle:2006vb}.

\textbf{Seesaw basics~\cite{schechter:1980gr}}

The most general seesaw is described in terms of the SM gauge
structure. Most of the low energy phenomenology, such as that of
neutrino oscillations, is blind to the details of the underlying
seesaw theory at high energies, e.~g. its gauge group, multiplet
content or the nature of B-L. The full seesaw mass matrix including
the SU(2) triplet (type-II) terms was first given
in~\cite{schechter:1980gr} and reads
\begin{equation}
\label{ss-matrix-0} {\mathcal M_\nu} = \left(\begin{array}{cc}
    M_1 & D \\
    {D}^{T}   & M_2 \\ 
\end{array}\right) .
\end{equation}
in the basis of ``left'' and ``right'' neutrinos $\nu_{L}$ and
$\nu^{c}_{L}$. Here we use the original notation of
reference~\cite{schechter:1980gr}, where the ``Dirac'' entry is
proportional to $\vev{\Phi}$, the $M_1$ comes from a triplet vev, and
$M_2$ is a gauge singlet.  The particular case $M_1=0$ was first
mentioned in Ref.~\cite{Minkowski:1977sc} and, subsequently, in
\cite{schechter:1980gr} and \cite{Orloff:2005nu,Lazarides:1980nt}.

Note that the matrix ${\mathcal M_\nu}$ is complex, so are its Yukawa
coupling sub-matrices \(D\), \(M_1\) and \(M_2\), the last two
symmetric, by the Pauli principle. It is diagonalized by a unitary
transformation \(U_\nu\),
\begin{equation}
\label{eq:light-nu}
   \nu_i = \sum_{a=1}^{6}(U_\nu)_{ia} n_a ,
\end{equation}
so that
\begin{equation}
   U_\nu^T {\mathcal M_\nu} U_\nu = \mathrm{diag}(m_i,M_i) .
\end{equation}
This yields 6 mass eigenstates, including the 3 light neutrinos with
masses \(m_i\), and 3 two-component heavy leptons of masses \(M_i\).
The light neutrino mass states \(\nu_i\) are given in terms of the
flavour eigenstates via eq.~(\ref{eq:light-nu}).
The effective light neutrino mass reads,
\begin{equation}
  \label{eq:ss-formula0}
  m_{\nu} \simeq M_1 - D {M_2}^{-1} {D}^T . 
\end{equation}
The smallness of light neutrino masses comes from the hierarchy $M_2
\gg D \gg M_1$.  A dynamical understanding of this hierarchy is
obtained in schemes where lepton number symmetry is broken
spontaneously, either with gauged or ungauged B-L.

\textbf{Simplest seesaw dynamics~\cite{schechter:1982cv}}

The simplest seesaw is based on the \321 gauge group with ungauged
lepton number.  The mass terms in eq.~(\ref{ss-matrix-0}) are given by
triplet, doublet and singlet vevs, respectively,
as~\cite{schechter:1982cv}
\begin{equation}  
\label{ss-matrix-123} {\mathcal M_\nu} = \left(\begin{array}{cc}
    Y_3 v_3 & Y_\nu \vev{\Phi} \\
    {Y_\nu}^{T} \vev{\Phi}  & Y_1 v_1 \\
\end{array}\right) 
\end{equation}
As already mentioned, \(Y_\nu\), \(Y_3\) and \(Y_1\) are complex.
Neutrino masses arise either by heavy \321 singlet ``right-handed''
neutrino exchange (type I) or by the small effective triplet vev (type
II), as illustrated in Fig.~\ref{fig:seesaw}.  The effective light
neutrino mass is easily obtained from Eq.~(\ref{eq:ss-formula0}) and
its diagonalization matrices containing the CP phases relevant in
leptogenesis (see below) are given explicitly as a matrix perturbation
series expansion in $D M_2^{-1}$~\cite{schechter:1982cv}.

Since lepton number is ungauged, there is a physical Goldstone boson
associated with its spontaneous breakdown, the majoron. Its profile
can be determined just by analysing the symmetry properties of the
scalar potential (not its detailed form) which dynamically determines
the vevs appearing in
Eq.~(\ref{ss-matrix-123})~\cite{schechter:1982cv}. These obey a simple
hierarchy $$v_1 \gg v_2 \gg v_3$$ with a vev seesaw relation of the
type $v_3 v_1 \sim {v_2}^2$ where $v_2 \equiv \vev{\Phi}$ denotes the
SM Higgs doublet vev, fixed by the W-boson mass.  This hierarchy
implies that the triplet vev $v_3 \to 0$ as the singlet vev $v_1$
grows and hence the type-II term is also suppressed.  This model
provides the first realization of seesaw that gives a dynamical
understanding of the smallness of both type-I and type-II terms.

\textbf{Left-right symmetric seesaw~\cite{Minkowski:1977sc,Orloff:2005nu}}
 
This is a more symmetric (less general) version of the seesaw, where
lepton number (B-L) is gauged. For example, it can be realized either
in terms of \10 or its \lr
subgroup~\cite{Minkowski:1977sc,Orloff:2005nu}.
In \10 each matter generation is assigned to a {\bf 16} (spinorial) so
that the {\bf 16} . {\bf 16} .  {\bf 10} and {\bf 16} . {\bf 126} .
{\bf 16} couplings generate all entries of the seesaw matrix in
Eq.~(\ref{eq:ss-formula0}) where \(Y_L\) and \(Y_R\) denote the
Yukawas of the {\bf 126} of \10, whose vevs $\vev{\Delta_{L,R}}$ give
rise to the Majorana terms. They correspond to \(Y_1\) and \(Y_3\) of
the simplest seesaw model.  On the other hand ${Y_\nu}$ denotes the
{\bf 16} . {\bf 16} .  {\bf 10} Dirac Yukawa coupling.
The diagonalization can be worked out as in the simplest case. With
obvious changes, e.~g. $v_1 \to \vev{\Delta_R}$ and $v_3 \to
\vev{\Delta_L}$, the explicit formulas for the \(6\times6\) unitary
diagonalizing matrix \(U\) given in Ref.~\cite{schechter:1982cv} also
hold.

The only important difference with respect to the previous case is the
absence of the majoron, now absorbed as the longitudinal mode of the
gauge boson coupled to B-L, which picks up a huge mass.  The smallness
of neutrino masses gets correlated to the observed maximality of
parity violation in low-energy weak interactions, a connection which
is as elegant as phenomenologically irrelevant, given the large value
of the B-L scale required both to fit the neutrino masses, as well as
to unify the gauge couplings.

 \textbf{Extended seesaw}

 In any gauge theory one can add any number of (anomaly-free) gauge
 singlet leptons~\cite{schechter:1980gr}. For example, in \10 and \e6
 one may add leptons outside the {\bf 16} or the {\bf 27},
 respectively.  Some of these extended seesaw
 schemes~\cite{mohapatra:1986bd} are motivated by string
 theories~\cite{Witten:1985xc}. New features emerge when the seesaw is
 realized non-minimally.  Recent examples are
 type-III~\cite{Akhmedov:1995vm,Barr:2005ss,Fukuyama:2005gg} and the
 \10 seesaw mechanism with low B-L scale in
 Ref.~\cite{Malinsky:2005bi}. For a brief review see
 Ref.~\cite{Valle:2006vb}.

\subsection{Low-scale models}
\label{sec:bottom-up-scenario}

There are many models of neutrino mass where the dim-5 operator is
induced from physics at low scales, TeV or less.  The smallness of its
strength comes then from loop and Yukawa couplings suppression and by
small lepton number violating parameters that appear in its numerator,
instead of its denominator. Here is an example.

{\bf Inverse seesaw  \cite{mohapatra:1986bd}}

It has the same mass matrix as the double seesaw
model~\cite{Valle:2006vb}, except that the basic L-violating scale
\(\mu\) is taken very small, e.~g.  \(\mu \ll Y_\nu \vev{\phi} \ll
M\)~\cite{mohapatra:1986bd}.  As a result neutrino masses vanish as
\(\mu\ \to 0\),
$$m_\nu =  {\vev{\Phi}^2} Y_\nu^T { M^{T}}^{-1} \mu M^{-1}  Y_\nu ,$$
opposite  to what happens in minimal seesaw.
The entry \(\mu\) may be proportional to the vev of an SU(2) singlet
scalar, in which case spontaneous B-L violation leads to the existence
of a majoron~\cite{gonzalez-garcia:1989rw}, implying a new phase
transition after the electroweak transition.
Since all particles are at the TeV scale, there are possibly testable
phenomenological implications, including \lfv in muon and tau
decays~\cite{Deppisch:2004fa}.

The model is ``natural'' in t'Hooft's sense~\cite{'tHooft:1979bh}:
{\it ``an otherwise arbitrary parameter may be taken as small when the
   Lagrangean symmetry increases by having it vanish''}.

{\bf Radiative models}

\begin{figure}[h] \centering
    \includegraphics[height=3cm,width=.45\linewidth]{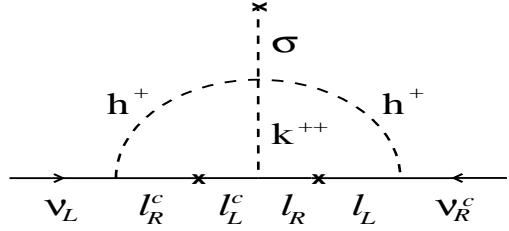}
    \caption{\label{fig:neumass} 
    Two-loop origin for neutrino mass~\cite{babu:1988ki,Peltoniemi:1993pd}.}
\end{figure}

Neutrino masses may be induced by calculable loop
corrections~\cite{zee:1980ai,babu:1988ki} as illustrated in
Fig.~\ref{fig:neumass}. For example in the the two-loop model one has,
up to a logarithmic factor,
\begin{equation}
   \label{eq:babu}
{\mathcal M_\nu} \sim \lambda_0 \left(\frac{1}{16\pi^2}\right)^2 
f Y_l h Y_l f^T \frac{\vev{\Phi}^2}{(m_k)^2} \vev{\sigma}
 \end{equation}
 in the limit where the doubly-charged scalar $k$ is much heavier than
 the singly charged one. Here $l$ denotes a charged lepton, $f$ and
 $h$ are their Yukawa coupling matrices and $Y_l$ denotes the SM Higgs
 Yukawa couplings to charged leptons. Here $\vev{\sigma}$ denotes an
 \321 singlet vev used in Ref.~\cite{Peltoniemi:1993pd}.  Clearly,
 even if the proportionality factor $\lambda_0$ is large, the neutrino
 mass is suppressed by the presence of a product of five small Yukawas
 and the appearance of the two-loop factor.

{\bf Supersymmetry and neutrino mass}

\begin{figure}[h] \centering
    \includegraphics[height=3cm,width=.5\linewidth]{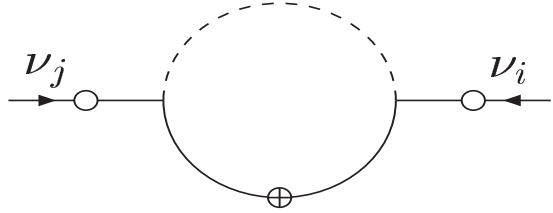}
    \caption{\label{fig:rpsusy} Loop-induced solar scale in
      RMSSM~\cite{Hirsch:2000ef}; open blobs denote $\Delta L=1$
      insertions.}
\end{figure}

The intrinsically supersymmetric way to break lepton number is to
break the so-called R parity. This may happen spontaneously, driven by
a nonzero vev of an \321 singlet
sneutrino~\cite{Masiero:1990uj,romao:1992vu,romao:1997xf}, leading to
an effective model with bilinear violation of R
parity~\cite{Diaz:1998xc}. This provides the minimal way to add
neutrino masses to the MSSM, we call it RMSSM~\cite{Hirsch:2004he}, to
stress that it serves as a reference model. Neutrino mass generation
is hybrid, with one scale generated at tree level and the other
induced by ``calculable'' radiative corrections~\cite{Hirsch:2000ef}.
The neutrino mass spectrum is typically ``normal hierarchy''-type,
with the atmospheric scale generated at the tree level and the solar
mass scale arising from calculable loops, as in Fig.~\ref{fig:rpsusy}.
The general form of the expression is quite involved but the
approximation
\begin{equation}
   \label{eq:bilinear}
{\mathcal M_\nu} \sim \left(\frac{1}{16\pi^2}\right) {\vev{\Phi}^2} \frac{A}{m_0} Y_d Y_d 
 \end{equation}
 (where $A$ denotes the trilinear soft supersymmetry breaking
 coupling) holds in some regions of parameters.

\subsection{Predicting  neutrino masses and mixing}
\label{sec:pred-neutr-mixing}

Currently five of the basic parameters of the lepton sector are probed
in neutrino oscillation studies.  Data points towards a well defined
pattern of neutrino mixing angles, quite distinct from that of quarks,
and difficult to account for in unified schemes where quarks and
leptons are related. There seems to be an intriguing complementarity
between quark and lepton mixing
angles~\cite{Raidal:2004iw,minakata-2004-70,Ferrandis:2004vp,Dighe:2006zk}.

%%%

There have been many papers trying to understand the values of the
leptonic mixing angles from underlying symmetries.  Of course, this is
part of the the problem of predicting quark and lepton mixings, a
defying challenge for model-builders.

Harrison, Perkins \& Scott have suggested~\cite{Harrison:2002kp} that
at high scales the neutrino mixing angles are given by,
\begin{align}
\label{eq:hps}
\tan^2\theta_{\Atm}&=\tan^2\theta_{23}^0=1\\ \nonumber
\sin^2\theta_{\textrm{Chooz}}&=\sin^2\theta_{13}^0=0\\
\tan^2\theta_{\Sol}&=\tan^2\theta_{12}^0=0.5 .\nonumber
\end{align}
Such pattern~\cite{Harrison:2002er} could result from some flavour
symmetry.  Its predictions should be corrected by renormalization
group
evolution~\cite{Altarelli:2005yp,Hirsch:2006je,Altarelli:2004za}.

Here I consider a specific idea to predict neutrino masses and mixing
angles: that neutrino masses arise from a common seed at some
``neutrino mass unification'' scale $M_X$~\cite{chankowski:2000fp},
very similar to the merging of the SM gauge coupling constants at high
energies due to supersymmetry~\cite{amaldi:1991cn}.
Although in its simplest form this idea is now inconsistent (at least
if CP is conserved) with the observed value of the solar mixing angle
$\theta_{12}$, there is an alternative realization in terms of an
$A_4$ flavour symmetry which is both viable and
predictive~\cite{babu:2002dz}. Starting from three-fold degeneracy of
the neutrino masses at the seesaw scale, the model predicts maximal
atmospheric angle and vanishing $\theta_{13}$,
$$\theta_{23}=\pi/4~~~\rm{and}~~~\theta_{13}=0\:.$$ 
Although the solar angle $\theta_{12}$ is unpredicted, one
expects~\footnote{There have been realizations of the $A_4$ symmetry
  that also predict the solar angle, e.~g.
  Ref.~\cite{Hirsch:2005mc}.}
$$\theta_{12}={\cal O}(1).$$ 
If CP is violated $\theta_{13}$ becomes arbitrary and the Dirac phase
is maximal~\cite{Grimus:2003yn}.  One can show that lepton and slepton
mixings are closely related and that there must exist at least one
slepton below 200 GeV, which can be produced at the LHC. The absolute
Majorana neutrino mass scale $m_0 \geq 0.3$ eV ensures that the model
will be probed by future cosmological tests and $\beta\beta_{0\nu}$
searches.  Rates for lepton flavour violating processes $l_j \to \l_i
+ \gamma$ typically lie in the range of sensitivity of coming
experiments, with BR$(\mu \to e \gamma) \gsim 10^{-15}$ and BR$(\tau
\to \mu \gamma) > 10^{-9}$.

\subsection{Absolute scale of neutrino mass and \znbb}
\label{sec:neutr-double-beta}

\begin{figure}[h]
  \centering
\includegraphics[width=6cm,height=4cm]{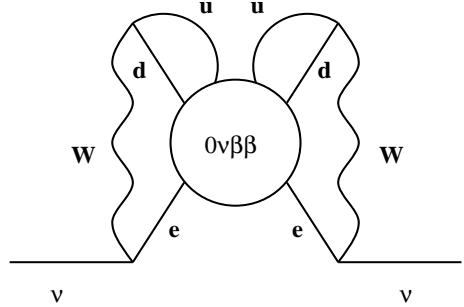}
\caption{Neutrinoless double beta decay and Majorana mass are
  equivalent~\cite{Schechter:1982bd}.}
 \label{fig:bbox}
\end{figure}

Neutrino oscillations are blind to whether neutrinos are Dirac or
Majorana. Lepton number violating processes, such as \znbb and
neutrino transition electromagnetic
moments~\cite{schechter:1981hw,Wolfenstein:1981rk}
\cite{pal:1982rm,kayser:1982br} probe the basic nature of neutrinos.
Neutrinoless double beta decay offers the best hope.  Its significance
stems from the fact that, in a gauge theory, irrespective of the
mechanism that induces \znbb, it necessarily implies a Majorana
neutrino mass~\cite{Schechter:1982bd}, as illustrated in Fig.
\ref{fig:bbox}.
Thus it is a basic issue. Quantitative implications of the
``black-box'' argument are model-dependent, but the theorem itself
holds in any ``natural'' gauge theory (for a recent discussion see
\cite{Hirsch:2006yk}).

%%%% 

\znbb will test absolute neutrino masses, inaccessible in neutrino
oscillations, and also complement direct information that will become
available from high sensitivity beta decay
studies~\cite{Drexlin:2005zt}, as well as cosmic microwave background
and large scale structure
observations~\cite{Lesgourgues:2006nd,Hannestad:2006zg,Fogli:2004as}.

The oscillation signal implies that \znbb must be induced by the
exchange of light Majorana neutrinos, through the so-called
"mass-mechanism". The corresponding amplitude is sensitive both to the
absolute scale of neutrino mass, and to Majorana
phases~\cite{schechter:1980gr}, neither of which can be probed in
oscillations~\cite{bilenky:1980cx,Schechter:1981gk}.
Taking into account current neutrino oscillation
parameters~\cite{Maltoni:2004ei} and state-of-the-art nuclear matrix
elements~\cite{Bilenky:2004wn} one can determine the average mass
parameter $\meff$ characterizing the neutrino exchange contribution to
\znbb, as shown in Fig. 10 of Ref.~\cite{Valle:2006vb}.
Models with quasi-degenerate
neutrinos~\cite{babu:2002dz}~\cite{caldwell:1993kn}~\cite{ioannisian:1994nx}
give the largest \znbb signal. For models with normal hierarchy there
is in general no lower bound on $\meff$ since there can be a
destructive interference amongst the neutrino amplitudes (for an
exception, see Ref.~\cite{Hirsch:2005mc}; in that specific model a
lower bound on $\meff$ exists, which depends, as expected, on the
value of the Majorana CP violating phase $\phi_1$).  In contrast, the
inverted neutrino mass hierarchy implies a ``lower'' bound for the
\znbb amplitude.

The best current limit on $\meff$ comes from the Heidelberg-Moscow
experiment. There is also a claim made in
Ref.~\cite{Klapdor-Kleingrothaus:2004wj} (see
also~\cite{Aalseth:2002dt}) which will be important to confirm or
refute in future experiments. GERDA will provide an independent check
of this claim~\cite{Aalseth:2002rf}. SuperNEMO, CUORE, EXO, MAJORANA
and possibly other experiments will further extend the sensitivity of
current \znbb searches~\cite{dbd06}.

\subsection{Other phenomena}
\label{sec:other-phenomena}

Besides oscillations and \znbb neutrino masses may have other
phenomenological manifestations. Here I summarize a few.

\begin{itemize}
\item {\bf \lfv} Now that \lfv has been shown to occur in neutrino
  propagation it is natural to expect that it may show up elsewhere.
  Indeed, it is expected to occur in seesaw-type schemes of neutrino
  mass, either from neutral heavy lepton
  exchange~\cite{Bernabeu:1987gr,gonzalez-garcia:1992be,Ilakovac:1994kj}
  of through supersymmetric
  contributions~\cite{Hall:1985dx,borzumati:1986qx,casas:2001sr,Antusch:2006vw}.
  Note that since flavor and CP violation can occur in the massless
  neutrino limit, the allowed rates are unsuppressed by the smallness
  of neutrino
  masses~\cite{Bernabeu:1987gr,gonzalez-garcia:1992be,branco:1989bn,rius:1990gk}.
  In the extended seesaw scheme~\cite{mohapatra:1986bd} one can
  understand the interplay of both types of contributions. It is
  shown~\cite{Deppisch:2004fa} that \(Br(\mu\to e\gamma)\) and the
  nuclear $\mu^--e^-$ conversion rates lie within planned
  sensitivities of future experiments such as
  PRISM~\cite{Kuno:2000kd}.
\item {\bf TeV neutral heavy leptons} Extended seesaw models like the
  inverse seesaw may contain quasi-Dirac neutral heavy leptons around
  TeV or so, that may be directly produced at
  accelerators~\cite{Dittmar:1990yg}.
\item {\bf majoron-emitting neutrino decays} If neutrino masses arise
  from a spontaneous breaking of global lepton number the associated
  Goldstone boson (majoron) may lead to neutrino decays
  \cite{schechter:1982cv}. Although these are rather slow, they may be
  astrophysically relevant and lead to interesting signals
  \cite{kachelriess:2000qc} at underground detectors.
\item {\bf TeV new gauge boson coupled to lepton number} If neutrino
  masses arise from spontaneous breaking of gauged lepton number
  ~\cite{valle:1987sq,Malinsky:2005bi}, there will exist a light new
  neutral gauge boson, \ZP that could be detected in searches for
  Drell-Yan processes at the LHC.
\item {\bf invisible Higgs boson decays} In low-scale models of
  neutrino mass with spontaneous breaking of global lepton number the
  majoron can lead to an invisible Higgs boson
  decays~\cite{Joshipura:1993hp,romao:1992zx,Hirsch:2004rw,Bazzocchi:2006vn}.
\begin{equation}
  \label{eq:JJ}
  H \to JJ
\end{equation}
where $J$ is the majoron.  The latter is experimentally detectable as
missing energy or transverse momentum associated to the
Higgs~\cite{deCampos:1997bg,Abdallah:2003ry}, a signal that must be
taken into account when designing Higgs boson search strategies at
future collider experiments.  This shows that, although neutrino
masses are small, the neutrino mass generation may have very important
implications for the mechanism of electroweak symmetry breaking.
\item {\bf Reconstructing neutrino mixing at accelerators} Low-scale
  models of neutrino mass offer the tantalizing possibility of
  reconstructing neutrino mixing at high energy accelerators, like the
  LHC and the ILC.  A remarkable example is provided by models where
  supersymmetry is the origin of neutrino mass~\cite{Hirsch:2004he}.
  A general feature of these models is that, unprotected by any
  symmetry, the lightest supersymmetric particle (LSP) is unstable. In
  order to reproduce the masses indicated by current neutrino
  oscillation data, the LSP is expected to decay inside the
  detector~\cite{Hirsch:2000ef}~\cite{deCampos:2005ri}.  More
  strikingly, LSP decay properties correlate with the neutrino mixing
  angles. For example, if the LSP is the lightest neutralino, it
  should have the same decay rate into muons and taus, since the
  observed atmospheric angle is close to
  $\pi/4$~\cite{Porod:2000hv,romao:1999up,mukhopadhyaya:1998xj}.
  Such correlations hold irrespective of which supersymmetric particle
  is the LSP~\cite{Hirsch:2003fe} and constitute a smoking gun
  signature of this proposal that will be tested at upcoming
  accelerators.
\end{itemize}

\subsection{Thermal leptogenesis}
\label{sec:thermal-leptogenesis}

\begin{figure}[h]
\centering
\includegraphics[clip,height=3.2cm,width=0.8\linewidth]{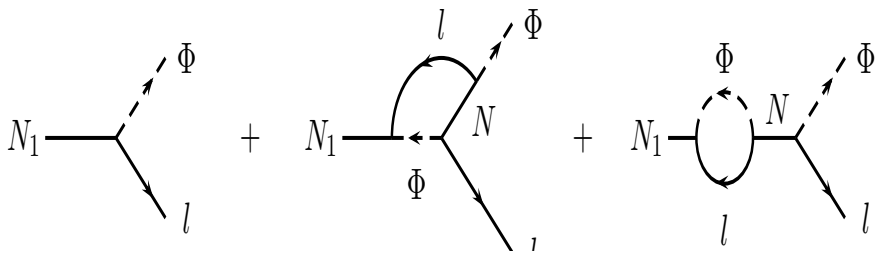}
\caption{Diagrams contributing to  leptogenesis.}
     \label{fig:lep-g}
\end{figure}

It has long been noted~\cite{Fukugita:1986hr} that seesaw models open
an attractive possibility of accounting for the observed cosmological
matter-antimatter asymmetry in the Universe through
leptogenesis~\cite{Buchmuller:2005eh}.
In this picture the decays of the heavy ``right-handed'' neutrinos
present in the seesaw play a crucial role.
These take place through diagrams in Fig.~\ref{fig:lep-g}. In order to
induce successful leptogenesis the decay must happen before the
electroweak phase transition~\cite{kuzmin:1985mm} and must also happen
out-of-equilibrium, i.~e. the decay rate must be less than the Hubble
expansion rate at that epoch. Another crucial ingredient is CP
violation in the lepton sector.
The lepton (or B-L) asymmetry thus produced then gets converted,
through sphaleron processes, into the observed baryon asymmetry.

In seesaw-type schemes the high temperature needed for leptogenesis
leads to an overproduction of gravitinos, which destroys the standard
predictions of Big Bang Nucleosynthesis (BBN).
In minimal supergravity models, with $m_{3/2} \sim$ 100 GeV to 10 TeV
gravitinos are not stable, decaying during or after BBN. Their rate of
production can be so large that subsequent gravitino decays completely
change the standard BBN scenario.
To prevent such ``gravitino crisis'' one requires an upper bound on
the reheating temperature $T_R$ after inflation, since the abundance
of gravitinos is proportional to $T_R$. This leads to a stringent
upper bound~\cite{Kawasaki:2004qu}, which is in conflict with the
temperature required for leptogenesis, $T_R > 2 \times 10^9$
GeV~\cite{Buchmuller:2004nz}. One way to cure this
conflict~\cite{Farzan:2005ez} is to add a small R-parity violating
$\lambda_i \hat{\nu^c}_i \hat{H}_u \hat{H}_d$ term in the
superpotential, where $\hat{\nu^c}_i$ are right-handed neutrino
supermultiplets. One can show that in the presence of this term, the
produced lepton-antilepton asymmetry can be enhanced.
An alternative suggestion~\cite{Hirsch:2006ft} was made in the context
of extended supersymmetric seesaw schemes. It was shown in this case
that leptogenesis can occur at the TeV scale through the decay of a
new singlet, thereby avoiding the gravitino crisis. Washout of the
asymmetry is effectively suppressed by the absence of direct couplings
of the singlet to leptons.

\vspace{.3cm}

{\em Acknowledgments:} 

I thank the organizers for hospitality. This work was supported by
Spanish grants FPA2005-01269, European commission RTN Contract
MRTN-CT-2004-503369 and ILIAS/N6 Contract RII3-CT-2004-506222.

\def\baselinestretch{1}

%\bibliography{sample}
%\bibliographystyle{h-physrev4}
%\bibliography{lgenesis-ref,valle-ref,nu-rev06}
%\bibliography{nu-rev06}

\end{document}